\documentclass[10pt,twocolumn,letterpaper]{article}

\usepackage{wacv}
\makeatletter
\@namedef{ver@everyshi.sty}{}
\makeatother
\usepackage{tikz}
\usepackage{times}
\usepackage{epsfig}
\usepackage{multicol}
\usepackage{multirow}
\usepackage{graphicx}
\usepackage{booktabs}
\usepackage[hyphens]{url}
\PassOptionsToPackage{hyphens}{url}
\usepackage{hyperref}
\usepackage{amsmath}
\usepackage{latexsym}
\usepackage{amssymb}
\usepackage{float}
\usepackage{enumitem}
\usepackage{pgfplots}
\usepackage{cite}
\usepackage{subcaption}
\wacvfinalcopy % *** Uncomment this line for the final submission

\def\wacvPaperID{1074} % *** Enter the wacv Paper ID here

\ifwacvfinal\fi
\DeclareUnicodeCharacter{2212}{-}
\begin{document}

%%%%%%%%% TITLE
\title{A Petri Dish for Histopathology Image Analysis}

% Authors at the same institution
%\author{First Author \hspace{2cm} Second Author \\
% Authors at different institutions
\author{Jerry Wei$^{1}$, Arief Suriawinata$^{2}$, Bing Ren$^{2}$, Xiaoying Liu$^{2}$, Mikhail Lisovsky$^{2}$,\\  
Louis Vaickus$^{2}$, Charles Brown$^{2}$, Michael Baker$^{2}$, Naofumi Tomita$^{1}$,\\
\vspace{1mm}
Lorenzo Torresani$^{1}$, Jason Wei$^{1}$, Saeed Hassanpour$^{1\dagger}$\\
\newline
\vspace{5mm}
$^{1}$Dartmouth College $^{2}$Dartmouth-Hitchcock Medical Center\\

\vspace{3mm}
$^\dagger$\texttt{saeed.hassanpour@dartmouth.edu}
\vspace{-4mm}
}

\maketitle
\ifwacvfinal\thispagestyle{empty}\fi

\vspace{-4mm}
\begin{abstract}
\vspace{-5mm}
With the rise of deep learning, there has been increased interest in using neural networks for histopathology image analysis, a field that investigates the properties of biopsy or resected specimens traditionally manually examined under a microscope by pathologists. 
However, challenges such as limited data, costly annotation, and processing high-resolution and variable-size images make it difficult to quickly iterate over model designs.

Throughout scientific history, many significant research directions have leveraged small-scale experimental setups as \textbf{petri dishes} to efficiently evaluate exploratory ideas. 
In this paper, we introduce a \textbf{m}inimalist \textbf{h}istopathology image analysis dataset (\textbf{MHIST}), an analogous petri dish for histopathology image analysis. 
MHIST is a binary classification dataset of 3,152 fixed-size images of colorectal polyps, each with a gold-standard label determined by the majority vote of seven board-certified gastrointestinal pathologists and annotator agreement level. 
MHIST occupies less than 400 MB of disk space, and a ResNet-18 baseline can be trained to convergence on MHIST in just 6 minutes using 3.5 GB of memory on a NVIDIA RTX 3090. 
As example use cases, we use MHIST to study natural questions such as how dataset size, network depth, transfer learning, and high-disagreement examples affect model performance.

By introducing MHIST, we hope to not only help facilitate the work of current histopathology imaging researchers, but also make the field more-accessible to the general community. 
Our dataset is available at \url{https://bmirds.github.io/MHIST}.

\end{abstract}

\begin{figure}[ht]
    \centering
    \includegraphics[width=0.9\linewidth]{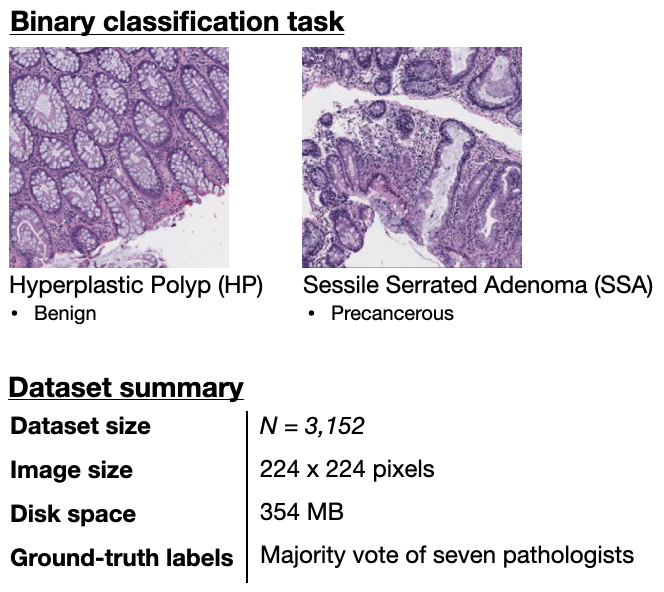}
    \vspace{-2mm}
    \caption{Key features of our \textbf{m}inimalist \textbf{hist}opathology image analysis dataset (MHIST).}
    \vspace{-3mm}
    \label{fig:pull_figure}
\end{figure}

\vspace{-7mm}
\section{Introduction}
\vspace{-2mm}
Scientific research has aimed to study and build our understanding of the world, and although many problems initially seemed too ambitious, they were ultimately surmounted.
In these quests, a winning approach has often been to break down large ideas into smaller components, learn about these components through experiments that can be iterated on quickly, and then validate or translate those ideas into large-scale applications.
For example, in the Human Genome Project (which helped us understand much of what we know now about human genetics), many fundamental discoveries resulted from \textit{petri dish} experiments---small setups that saved time, energy, and money---on simpler organisms. 
In particular, the Drosophila fruit fly, an organism that is inexpensive to culture, has short life cycles, produces large numbers of embryos, and can be easily genetically modified, has been used in biomedical research for over a century to study a broad range of phenomena \cite{JENNINGS2011190}.

%some motivation for petri dish analogy
In deep learning, we have our own set of petri dishes in the form of benchmark datasets, of which MNIST \cite{LeCun1998} is one of the most popular.
Comprising the straightforward problem of classifying handwritten digits in 28 by 28 pixel images, MNIST is easily accessible, and training a strong classifier on it has become a simple task with today's tools.
Because it is so easy to evaluate models on MNIST, it has served as the exploratory environment for many ideas that were then validated on large scale datasets or implemented in end-to-end applications. 
For example, many well-known concepts such as convolutional neural networks, generative adversarial networks \cite{Goodfellow2014}, and the Adam optimization algorithm \cite{Kingma2014} were initially validated on MNIST. 

%talk about how nothing exists for histopath, and potentially why
In the field of histopathology image analysis, however, no such classic dataset currently exists due to many potential reasons.
To start, most health institutions do not have the technology nor the capacity to scan histopathology slides at the scale needed to create a reasonably-sized dataset.
Even for institutions that are able to collect data, a barrage of complex data processing and annotation decisions falls upon the aspiring researcher, as histopathology images are large and difficult to process, and data annotation requires the valuable time of trained pathologists.
Finally, even after data is processed and annotated, it can be challenging to obtain institutional review board (IRB) approval for releasing such datasets, and some institutions may wish to keep such datasets private. 
As a result of the inaccessibility of data, histopathology image analysis has remained on the fringes of computer vision research, with many popular image datasets dealing with domains where data collection and annotation are more straightforward.

%talk about what our dataset is and how it fills this gap in the research
To address these challenges that have plagued deep learning for histopathology image analysis since the beginning of the area, in this paper, we introduce \textbf{MHIST}: a \textbf{m}inimalist \textbf{hist}opathology image classification dataset.
MHIST is minimalist in that it comprises a straightforward binary classification task of fixed-size colorectal polyp images, a common and clinically-significant task in gastrointestinal pathology. 
MHIST contains 3,152 fixed-size images, each with a gold-standard label determined from the majority vote of seven board-certified gastrointestinal pathologists, that can be used to train a baseline model without additional data processing. 
By releasing this dataset publicly, we hope not only that current histopathology image researchers can build models faster, but also that general computer vision researchers looking to apply models to datasets other than classic benchmarks can easily explore the exciting area of histopathology image analysis. 
Our dataset is publicly available at \url{https://bmirds.github.io/MHIST} following completion of a simple dataset-use agreement form.

\vspace{-3mm}
\section{Background}
\vspace{-1mm}

Deep learning for medical image analysis has recently seen increased interest in analyzing histopathology images (large, high-resolution scans of histology slides that are typically examined under a microscope by pathologists) \cite{srinidhi2020deep}. 
To date, deep neural networks have already achieved pathologist-level performance on classifying diseases such as prostate cancer, breast cancer, lung cancer, and melanoma \cite{Arvaniti2018,Bulten2020,Hekler2019,Shah2017,Strom2019,Wei2019,Zhang2019}, demonstrating their large potential.
Despite these successes, histopathology image analysis has not seen the same level of popularity as analysis of other medical image types (e.g., radiology images or CT scans), likely because the nature of histopathology images creates a number of hurdles that make it challenging to directly apply mainstream computer vision methods. 
Below, we list some factors that can potentially impede the research workflow in histopathology image analysis:

\begin{itemize}
    \item \textbf{High-resolution, variable-size images.} 
    Because the disease patterns in histology slides can only occur in certain sections of the tissue and can only be detected at certain magnifications under the microscope, histopathology images are typically scanned at high resolution so that all potentially-relevant information is preserved. 
    This means that while each sample contains lots of data, storing these large, high-resolution images is nontrivial.
    For instance, the slides from a single patient in the CAMELYON17 challenge \cite{camelyon17} range from 2 GB to 18 GB in size, which is up to one-hundred times larger than the entire CIFAR-10 dataset. 
    Moreover, the size and aspect ratios of the slides can differ based on the shape of the specimen in question---sometimes, multiple large specimens are included in one slide, and so some scanned slides may be up to an order of magnitude larger than others.
    As deep neural networks typically require fixed-dimension inputs, preprocessing decisions such as what magnification to analyze the slides at and how to deal with variable-size inputs can be difficult to make.
    \item \textbf{Cost of annotation.}
    Whereas annotating data in deep learning has been simplified by services such as Mechanical Turk, there is no well-established service for annotating histopathology images, a process which requires substantial time from experienced pathologists who are often busy with clinical service. 
    Moreover, access to one or two pathologists is often inadequate because inter-annotator agreement is low to moderate for most tasks, and so annotations can be easily biased towards the personal tendencies of annotators.
    \item \textbf{Unclear annotation guidelines.}
    It is also unclear what type of annotation is needed for high-resolution whole-slide images, as a slide may be given a certain diagnosis based on a small portion of diseased tissue, but the overall diagnosis would not apply to the normal portions of the tissue.
    Researchers often opt to have pathologists draw bounding boxes and annotate areas with their respective histological characteristics, but this comes with substantial costs, both in training pathologists to use annotation software and in increased annotation time and effort.
    \item \textbf{Lack of data.}
    Even once these challenges are addressed, it is often the case that, due to slides being discarded as a result of poor quality or to remove classes that are too rare to include in the classification task, training data is relatively limited and the test set is not sufficiently large. 
    This makes it difficult to distinguish accurately between models, and models are therefore easily prone to overfitting. 
\end{itemize}

To mitigate these challenges of data collection and annotation, in this paper we introduce a minimalist histopathology dataset that will allow researchers to quickly train a histopathology image classification model without dealing with an avalanche of complex data processing and annotation decisions.
Our dataset focuses on the binary classification of colorectal polyps, a straightforward task that is common in a gastrointestinal pathologist's workflow. 
Instead of using whole-slide images, which are too large to directly train on for most academic researchers, our dataset consists only of 224 $\times$ 224 pixel image tiles of tissue; these images can be directly fed into standard computer vision models such as ResNet. 
Finally, for annotations, each patch in our dataset was directly classified by seven board-certified gastrointestinal pathologists and given a gold-standard label based on their majority vote.

Our dataset aims to serve as a petri dish for histopathology image analysis. 
That is, it represents a simple task that can be learned quickly, and it is easy to iterate over.
Our dataset allows researchers to, without dealing with the confounding factors that arise from the nature of histopathology images, quickly test inductive biases that can later be implemented in large-scale applications. 
We hope that our dataset will allow researchers to more-easily explore histopathology image analysis and that this can facilitate further research in the field as a whole.

\section{MHIST Dataset}

In the context of the challenges mentioned in the above section, MHIST has several notable features that we view favorably in a minimalist dataset:
\begin{enumerate}
    \item Straightforward binary classification task that is challenging and important.
    \item Adequate yet tractable number of examples: 2,175 training and 977 testing images.
    \item Fixed-size images of appropriate dimension for standard models.
    \item Gold-standard labels from the majority vote of seven pathologists, along with annotator agreement levels that can be used for more-specific model tuning.
\end{enumerate}

The rest of this section details the colorectal polyp classification task ($\S$\ref{subsec:cp-task}), data collection ($\S$\ref{subsec:data-collection}), and the data annotation process ($\S$\ref{subsec:data-annotation}). 

\subsection{Colorectal Polyp Classification Task}\label{subsec:cp-task}
Colorectal cancer is the second leading cause of cancer death in the United States, with an estimated 53,200 deaths in 2020 \cite{ColorectalCancerStats}.
As a result, colonoscopy is one of the most common cancer screening programs in the United States \cite{Rex2017}, and classification of colorectal polyps (growths inside the colon lining that can lead to colonic cancer if left untreated) is one of the highest-volume tasks in pathology.
Our task focuses on the clinically-important binary distinction between hyperplastic polyps (HPs) and sessile serrated adenomas (SSAs), a challenging problem with considerable inter-pathologist variability \cite{Abdeljawad2015,Farris2008,Glatz2007,Khalid2009,Wong2009}.
HPs are typically benign, while SSAs are precancerous lesions that can turn into cancer if left untreated and require sooner follow-up examinations \cite{HPvsSSA}.
Pathologically, HPs have a superficial serrated architecture and elongated crypts, whereas SSAs are characterized by broad-based crypts, often with complex structure and heavy serration \cite{Gurudu2010}.

\subsection{Data Collection}\label{subsec:data-collection}
For our dataset, we scanned 328 Formalin Fixed Paraffin-Embedded (FFPE) whole-slide images of colorectal polyps, which were originally diagnosed on the whole-slide level as hyperplastic polyps (HPs) or sessile serrated adenomas (SSAs), from patients at the Dartmouth-Hitchcock Medical Center.
These slides were scanned by an Aperio AT2 scanner at 40x resolution; to increase the field of view, we compress the slides with 8x magnification. 
From these 328 whole-slide images, we then extracted 3,152 image tiles (portions of size 224 $\times$ 224 pixels) representing diagnostically-relevant regions of interest for HPs or SSAs.
These images were shuffled and anonymized by removing all metadata such that no sensitive patient information was retrievable from any images. 
All images contain mostly tissue by area (as opposed to white space) and were confirmed by our pathologists to be high-quality with few artifacts. 
The use and release of our dataset was approved by Dartmouth-Hitchcock Health IRB.

\begin{table}[t!]
    \centering
    \begin{tabular}{l c c c}
        \toprule
         & Train & Test & Total\\
        \midrule
         HP & 1,545 & 617 & 2,162\\
         SSA & 630 & 360 & 990\\
         \midrule
         Total & 2,175 & 977 & 3,152\\
         \bottomrule
    \end{tabular}
    \caption{Number of images in our dataset's training and testing sets for each class. HP: hyperplastic polyp (benign), SSA: sessile serrated adenoma (precancerous).}
    \label{tab:train_test_split}
\end{table}

\begin{figure}[h]
    \centering
    \includegraphics[width=\linewidth]{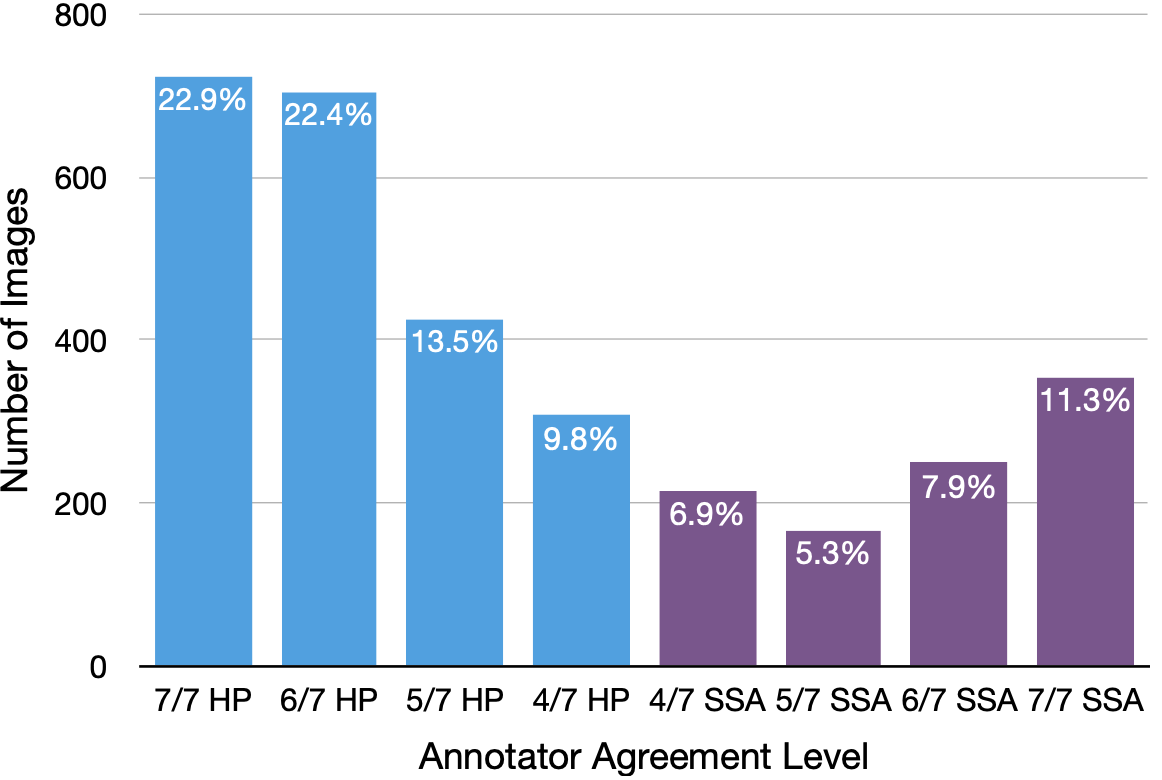}
    \caption{Distribution of annotator agreement levels for images in our dataset.}
    \vspace{-7mm}
    \label{fig:annotator_agreement_distribution}
\end{figure}

\subsection{Data Annotation}\label{subsec:data-annotation}
For data annotation, we worked with seven board-certified gastrointestinal pathologists at the Dartmouth-Hitchcock Medical Center.
Each pathologist individually and independently classified each image in our dataset as either HP or SSA based on the World Health Organization criteria from 2019 \cite{nagtegaal20202019}. 
After labels were collected for all images from all pathologists, the gold standard label for each image was assigned based on the majority vote of the seven individual labels, a common choice in literature \cite{Chilamkurthy2018,Gulshan2016,Irvin2019,Kanavati2020,Korbar2017,Sertel2008,Wang2019,Wei2019Celiac,Wei2020Difficulty,Zhou2019}. 
The distribution of each class in our dataset based on the gold standard labels of each image is shown in Table \ref{tab:train_test_split}.

In our dataset, the average percent agreement between each pair of annotators was 72.9\%, and each pathologist agreed with the majority vote an average of 83.2\% of the time. 
There is, notably, nontrivial disagreement between pathologists (approximately 16.7\% of images have 4/7 agreement), which corresponds with the difficulty of our colorectal polyp classification task.
The mean of the per-pathologist Cohen's $\kappa$ was 0.450, in the moderate range of 0.41-0.60. 
Although not directly comparable with prior work, a similar evaluation found a Cohen's $\kappa$ of 0.380 among four pathologists \cite{Wong2009}.
To facilitate research that might consider the annotator agreement of examples during training, for each image, we also provide the agreement level among our annotators (4/7, 5/7, 6/7, or 7/7). 
Figure \ref{fig:annotator_agreement_distribution} shows the distribution of agreement levels for our dataset.

\vspace{-4mm}
\section{Example Use Cases}
\vspace{-2mm}

In this section, we demonstrate example use cases of our dataset by investigating several natural questions that arise in histopathology image analysis.
Namely, how does network depth affect model performance ($\S$\ref{subsec:model-depth})?
How much does ImageNet pre-training help ($\S$\ref{subsec:transfer-learning})? 
Should examples with substantial annotator disagreement be included in training ($\S$\ref{subsec:high-disagreement})?
Moreover, we vary the size of the training set to gain insight on how the amount of available data interacts with each of the above factors. 

\begin{table}[t]
    \setlength{\tabcolsep}{5pt}
    \small
    \centering
    \begin{tabular}{c | c c c c}
        \toprule
        & \multicolumn{4}{c}{AUC (\%) on test set by training set size} \\
        ResNet & $n = 100$ & $n = 200$ & $n = 400$ & Full\\
        \midrule
        18 & \textbf{67.4 $\pm$ 3.1} & 73.6 $\pm$ 3.7 & \textbf{79.3 $\pm$ 2.3} & 84.5 $\pm$ 1.1\\
        34 & 64.1 $\pm$ 2.2 & \textbf{74.8 $\pm$ 3.1} & 78.0 $\pm$ 2.4 & \textbf{85.1 $\pm$ 0.7}\\
        50 & 64.7 $\pm$ 3.1 & 72.2 $\pm$ 2.8 & 76.6 $\pm$ 1.9 & 83.0 $\pm$ 0.6\\
        101 & 65.2 $\pm$ 5.6 & 73.2 $\pm$ 1.9 & 77.3 $\pm$ 0.9 & 83.2 $\pm$ 1.3\\
        152 & 62.3 $\pm$ 3.3 & 73.3 $\pm$ 1.2 & 77.5 $\pm$ 1.9 & 83.5 $\pm$ 0.8\\
        \bottomrule
    \end{tabular}
    \caption{Model performance for five different ResNet depths. Adding more layers to the model does not improve performance. $n$ indicates the number of images per class used for training. Means and standard deviations shown are for 10 random seeds.}
    \vspace{-4mm}
    \label{tab:network_depth}
\end{table}

\subsection{Experimental Setup}
For our experiments, we follow the DeepSlide code repository \cite{Wei2019} and use the ResNet architecture, a common choice for classifying histopathology images.
Specifically, for our default baseline, we use ResNet-18 and train our model for 100 epochs (well past convergence) using stochastic data augmentation with the Adam optimizer \cite{Kingma2014}, batch size of 32, initial learning rate of $1\times10^{-3}$, and learning rate decay factor of 0.91.

For more-robust evaluation, for each model we consider the five highest AUCs on the test set, which are evaluated at every epoch. 
We report the mean and standard deviation of these values calculated over 10 different random seeds.

Furthermore, we train our models with four different training set sizes: $n = 100$, $n = 200$, $n = 400$, and Full, where $n$ is the number of training images per class and Full is the entire training set.
To obtain subsets of the training set, we randomly sample $n$ random images for each class from the training set for each seed.
We keep our testing set fixed to ensure that models are evaluated equally.

\subsection{Network Depth}\label{subsec:model-depth}

We first study whether adding more layers to our model improves performance on our dataset.
Because deeper models take longer to train, identifying the smallest model that achieves the best performance allows for maximum accuracy with the least necessary training time.

We evaluate all five ResNet models proposed in \cite{He2015}---ResNet-18, ResNet-34, ResNet-50, ResNet-101, and ResNet-152---on our dataset, and all hyperparameters (e.g., number of epochs, batch size) are kept constant; we only change the model depth.

As shown in Table \ref{tab:network_depth}, adding more layers does not significantly improve performance.
Furthermore, adding model depth past ResNet-34 actually decreases performance, as models that are too deep will begin to overfit, especially for small training set sizes.
For example, when training with only 100 images per class, mean AUC decreases by 5.1\% when using ResNet-152 compared to ResNet-18.

We posit that increasing network depth does not improve performance on our dataset because our dataset is relatively small, and so deeper networks are unnecessary for the amount of information in our dataset.
Moreover, increasing network depth may increase overfitting on training data due to our dataset's small size.
Our results are consistent with findings presented by Benkendorf and Hawkins \cite{BENKENDORF2020}---deeper networks only perform better than shallow networks when trained with large sample sizes.

\vspace{-2mm}
\subsection{Transfer Learning}\label{subsec:transfer-learning}
\vspace{-2mm}

\begin{table}[t]
    \setlength{\tabcolsep}{3.5pt}
    \small
    \centering
    \begin{tabular}{c | c c c c}
        \toprule
        & \multicolumn{4}{c}{AUC (\%) on test set by training set size} \\
        Pretraining? & $n = 100$ & $n = 200$ & $n = 400$ & Full\\
        \midrule
        No & 67.4 $\pm$ 3.1 & 73.6 $\pm$ 3.7 & 79.3 $\pm$ 2.3 & 84.5 $\pm$ 1.1\\
        Yes & \textbf{83.7 $\pm$ 1.7} & \textbf{89.3 $\pm$ 1.8} & \textbf{92.4 $\pm$ 0.7} & \textbf{92.7 $\pm$ 0.4}\\
        \bottomrule
    \end{tabular}
    \caption{Using weights pretrained on ImageNet significantly improves the performance of ResNet-18 on our dataset. $n$ indicates the number of images per class used for training. Means and standard deviations shown are for 10 random seeds.}
    \vspace{-4mm}
    \label{tab:transfer_learning}
\end{table}

We also examine the usefulness of transfer learning for our dataset, as transfer learning can often be easily implemented into existing models, and so it is helpful to know whether or not it can improve performance.

Because deeper models do not achieve better performance on our dataset (as shown in Section \ref{subsec:model-depth}), we use ResNet-18 initialized with random weights as the baseline model for this experiment.
We compare our baseline with an identical model (i.e., all hyperparameters are congruent) that has been initialized with weights pretrained on ImageNet.

Table \ref{tab:transfer_learning} shows the results for our ResNet-18 model with and without pretraining. 
We find that ResNet-18 initialized with ImageNet pretrained weights significantly outperforms ResNet-18 initialized with random weights.
For example, our pretrained model's performance when trained with only 100 images per class is comparable to our baseline model's performance when trained with the full training dataset.
When both models are trained on the full training set, the pretrained model outperforms the baseline by 8.2\%, as measured by mean AUC.
These results indicate that, for our dataset, using pretrained weights can be extremely helpful for improving overall performance.

The large improvement from ImageNet pretraining is unlikely to result from our dataset having features expressed in ImageNet because ImageNet does not include histopathology images.
Instead, the improvement is, perhaps, explained by our dataset's small size, as ImageNet pretraining may help prevent the model from overfitting.
This would be consistent with Kornblith et al. \cite{Kornblith2018}, which found that performance improvements from ImageNet pretraining diminishes as dataset size increases.

\subsection{High-Disagreement Training Examples}\label{subsec:high-disagreement}
\vspace{-2mm}

\begin{table}[t!]
    \centering
    \begin{tabular}{l | c}
        \toprule
        Training Images Used & AUC (\%) on test set\\
        \midrule
        Very Easy Images Only & 79.9 $\pm$ 0.8\\
        Easy Images Only & 83.1 $\pm$ 0.6\\
        Very Easy + Easy Images & 84.6 $\pm$ 0.8\\
        Very Easy + Easy + Hard Images & \textbf{85.1 $\pm$ 0.8}\\
        All Images & 84.5 $\pm$ 1.1\\
        \bottomrule
    \end{tabular}
    \caption{Removing high-disagreement images during training may slightly improve performance. Means and standard deviations shown are for 10 random seeds.}
    \vspace{-4mm}
    \label{tab:disagreement_images}
\end{table}

As many datasets already contain annotator agreement data \cite{Chilamkurthy2018,Coudray2017,Bejnordi2017,Esteva2017,Ghorbani2019,Gulshan2016,Irvin2019,Kanavati2020,Korbar2017,Sertel2008,Wang2019,Wei2020,Zhou2019}, we also study whether there are certain ways of selecting examples based on their annotator agreement level that will maximize performance.
Examples with high annotator disagreement are, by definition, harder to classify, so they may not always contain features that are beneficial for training models.
For this reason, we focus primarily on whether training on only examples with higher annotator agreement will improve performance.

For our dataset, which was labeled by seven annotators, we partition our images into four discrete levels of difficulty: \textit{very easy} (7/7 agreement among annotators), \textit{easy} (6/7 agreement among annotators), \textit{hard} (5/7 agreement among annotators), and \textit{very hard} (4/7 agreement among annotators), following our prior work \cite{wei2020learn}.
We then train ResNet-18 models using different combinations of images selected based on difficulty: very easy images only; easy images only; very easy and easy images; and very easy, easy, and hard images.
For this experiment, we do not modify the dataset size like we did in Sections \ref{subsec:model-depth} and \ref{subsec:transfer-learning}, as selecting training images based on difficulty inherently changes the training set size.

As shown in Table \ref{tab:disagreement_images}, we find that excluding images with high annotator disagreement (i.e., hard and very hard images) during training achieves comparable performance to training with all images.
Using only very easy images or only easy images, however, does not match or exceed performance when training with all images.
We also find that training with all images except very hard images slightly outperforms training with all images.
One explanation for this is that very hard images, which only have 4/7 annotator agreement, could be too challenging to analyze accurately (even for expert humans), so their features might not be beneficial for training machine learning models either.

\begin{table*}[ht!]
    \centering
    \footnotesize
    \begin{tabular}{l | c c c c c c}
        \toprule
        Dataset & Task & Images & Image Type & Annotation Type & Annotators & Dataset Size\\
        \midrule
        MITOS (2012) \cite{Ciresan2013} & Mitosis Detection & 50 & High Power Fields & Pixel-Level & 2 Pathologists & $\sim$1 GB\\
        TUPAC16 \cite{VETA2019} & Mitosis Counting & 821 & Whole Slide Images & Regions of Interest & 1 Pathologist & $\sim$850 GB\\
        CAMELYON17 \cite{camelyon17} & Metastasis Detection & 1,000 & Whole Slide Images & Contour of Locations & 1 Pathologist & $\sim$2.3 TB\\
        PCam (2018) \cite{Veeling2018-qh} & Metastasis Detection & 327,680 & Fixed-Sized Images & Image-Wise & 1 Pathologist & $\sim$7 GB\\
        BACH (2018) \cite{BACH} & Breast Cancer Classification & 500 & Microscopy Images & Image-Wise & 2 Experts & $>$5 GB\\
        LYON19 \cite{LYON19} & Lymphocyte Detection & 83 & Whole Slide Images & Regions of Interest & 3 Analysts & $\sim$13 GB\\ 
        \midrule
        MHIST (Ours) & Colorectal Polyp Classification & 3,152 & Fixed-Sized Images & Image-Wise & 7 Pathologists & $\sim$333 MB\\
        \bottomrule
    \end{tabular}
    \vspace{-2mm}
    \caption{Comparison of well-known histopathology datasets. Our proposed dataset, MHIST, is advantageous due to its relatively small size (making it faster to obtain results) and its robust annotations.}
    \vspace{-3mm}
    \label{tab:comparison_of_datasets}
\end{table*}

\vspace{-1mm}
\section{Related Work}
\vspace{-1mm}
Due to the trend towards larger, more computationally-expensive models \cite{brown2020language}, as well as recent attention on the environmental considerations of training large models \cite{strubell2019energy}, the deep learning community has begun to question whether model development needs to occur at scale. 
In the machine learning field, two recent papers have brought attention to this idea.
Rawal et al.\ \cite{rawal2020synthetic} proposed a novel surrogate model for rapid architecture development, an artificial setting that predicts the ground-truth performance of architectural motifs. 
Greydanus \cite{greydanus2020scaling} proposed MNIST-1D, a low-computational alternative resource to MNIST that differentiates more clearly between models.
Our dataset falls within this direction and is heavily inspired by this work. 

In the histopathology image analysis domain, several datasets are currently available. 
Perhaps the two best-known datasets are CAMELYON17 \cite{camelyon17} and PCam \cite{Veeling2018-qh}. 
CAMELYON17 focuses on breast cancer metastasis detection in whole-slide images (WSIs) and includes a training set of 1,000 WSI with labeled locations.
CAMELYON17 is well-established, but because it contains WSIs (each taking up $>$1 GB), there is a large barrier to training an initial model, and it is unclear how to best pre-process the data to be compatible with current neural networks.

PCam is another well-known dataset that contains 327,680 images of size 96 $\times$ 96 pixels extracted from CAMELYON17.
While PCam is similar to our work in that it considers fixed-size images for binary classification, we note two key differences.
First, the annotations in PCam are derived from bounding-box annotations which were drawn by a student and then checked by a pathologist. 
For challenging tasks with high annotator disagreement, however, using only a single annotator can cause the model to learn specific tendencies of a single pathologist.
In our dataset, on the other hand, each image is directly classified by seven expert pathologists, and the gold standard is set as the majority vote of the seven labels, mitigating the potential biases that can arise from having only a single annotator.
Second, whereas PCam takes up around 7 GB of disk space, our dataset aims to be minimalist and is therefore an order of magnitude smaller, making it faster for researchers to obtain results and iterate over models.

In Table \ref{tab:comparison_of_datasets}, we compare our dataset with other previously-proposed histopathology image analysis datasets.
Our dataset is much smaller than other datasets, yet it still has enough examples to serve as a petri dish in that it can test models and return results quickly.
Additionally, our dataset has robust annotations in comparison to other histopathology datasets.
Datasets frequently only have one or two annotators, but MHIST is annotated by seven pathologists, making it the least influenced by biases that any singular annotator may have. 

\vspace{-2mm}
\section{Discussion}
\vspace{-2mm}
The inherent nature of histopathology image classification can create challenges for researchers looking to apply mainstream computer vision methods.
Histopathology images themselves are difficult to handle because they have high resolutions and are variable-sized, and accurately and efficiently annotating histopathology images is a nontrivial task.
Furthermore, being able to address these challenges does not guarantee a high-quality dataset, as histopathology images are difficult to acquire, and so data is often quite limited.
Based on a thorough analysis of these challenges, we have presented MHIST, a histopathology image classification dataset with a straightforward yet challenging binary classification task.
MHIST comprises a total of 3,152 fixed-size images that have already been preprocessed.
In addition to providing these images, we also include each image's gold standard label and annotator agreement.

Of possible limitations, our use of fixed-size images may not be the most-precise approach, as using whole-slide images directly would likely improve performance since whole-slide images contain much more information than fixed-size images.
Current computer vision models cannot train on whole-slide images, however, as a single whole-slide image can take up more than 10GB of space.
Thus, our dataset includes images of fixed-size, as this is appropriate for most standard computer vision models.
Another limitation is that our dataset does not include any demographic information about patients nor any information regarding the size and location of the polyp (data that is often used in clinical classification).
Our dataset contains purely image data and is limited in this fashion.

In this paper, we aim to have provided a dataset that can serve as a petri dish for histopathology image analysis.
We hope that researchers are able to use MHIST to test models on a smaller scale before being implemented in large-scale applications, and that our dataset will facilitate further research into deep learning methodologies for histopathology image analysis.

\clearpage
{\small
\bibliographystyle{splncs04}
\bibliography{egbib}
}

% \newpage
% \section*{Supplementary Materials}\label{apd:first}

\end{document}